\documentclass[conference]{IEEEtran}
\usepackage{booktabs}
\usepackage{makecell}
\usepackage{algorithm}
\usepackage{multirow}
\usepackage{array}
\usepackage{diagbox}
\usepackage{pifont}
\usepackage{cite}

\usepackage{graphicx}

\usepackage[cmex10]{amsmath}
\usepackage{amssymb}

\usepackage{algorithmicx}
\usepackage{algpseudocode}

\usepackage{array}

\usepackage{eqparbox}

\usepackage[caption=false,font=footnotesize]{subfig}

\usepackage{bm}

\usepackage{flushend}

\usepackage{cleveref}
\crefname{figure}{Fig.}{Figs.}
\crefname{section}{Section}{Sections}

\usepackage[dutch,english]{babel}

\begin{document}
%
\title{LINKs: Large Language Model Integrated Management for 6G Empowered Digital Twin NetworKs}


\author{Shufan Jiang, Bangyan Lin, Yue Wu\textsuperscript{*} and Yuan Gao}

\maketitle

\begingroup
\renewcommand{\thefootnote}{\relax}
\footnotetext{Shufan Jiang, Bangyan Lin and Yue Wu are with School of Information Science and Engineering, East China University of Science and Technology, China.}
\footnotetext{Yuan Gao is with Department of Electronic and Electrical Engineering, Shanghai University, China.}
\renewcommand{\thefootnote}{\fnsymbol{footnote}}
\footnotetext[1]{Corresponding to: \texttt{yuewu@ecust.edu.cn}}
\endgroup

\begin{abstract}
In the rapidly evolving landscape of digital twins (DT) and 6G networks, the integration of large language models (LLMs) presents a novel approach to network management. This paper explores the application of LLMs in managing 6G-empowered DT networks, with a focus on optimizing data retrieval and communication efficiency in smart city scenarios. The proposed framework leverages LLMs for intelligent DT problem analysis and radio resource management (RRM) in fully autonomous way without any manual intervention. Our proposed framework --- LINKs, builds up a lazy loading strategy which can minimize transmission delay by selectively retrieving the relevant data. Based on the data retrieval plan, LLMs transform the retrieval task into an numerical optimization problem and utilizing solvers to build an optimal RRM, ensuring efficient communication across the network. Simulation results demonstrate the performance improvements in data planning and network management, highlighting the potential of LLMs to enhance the integration of DT and 6G technologies. 

\end{abstract}

\section{Introduction}

\label{sec:intro}
The advent of DT has significantly transformed various industries by enabling the creation of virtual replicas of physical entities for simulation, monitoring, and prediction purposes. Concurrently, the development of 6G networks promises unprecedented improvements in network performance, enabling high-speed, low-latency, and highly reliable communications \cite{6G_Vision}. In the context of our study, DT is utilized to model the digitization of smart cities, while 6G technology is employed to enhance communication capabilities and global scheduling among device nodes defined under the concept of IoT. The integration of DT and 6G technologies opens new avenues for optimizing network management, strengthening predictive maintenance, enhancing security, and facilitating real-time monitoring and management within smart city systems \cite{6G_smart_city}.

In recent years, the rapid development of LLM, such as GPT series \cite{gpt4} and LLaMA series \cite{llama}, have shown remarkable capabilities in understanding real-world cases and generating human-level response, making them valuable tools for a wide range of tasks and applications, from natural language processing to data analysis \cite{bubeck2023sparksartificialgeneralintelligence}. However, the application of LLM in network management remains relatively unexplored. Some preliminary efforts have shown the potential of LLM to assist in network configuration and optimization\cite{LLM-network-opt}, but these approaches do not fully exploit the capabilities of LLMs for intelligent data retrieval and communication optimization within the context of DT and 6G networks.

Complex communication scenarios such as DT lead to the explosive growth of IoT devices and the complexity of network management. The dense deployment of devices leads to severe signal interference and competition for spectrum resources, which makes resource allocation more complex.

Under the circumstances, LLM has demonstrated its ability to effectively deal with complex network management through its capabilities of problem processing. \cite{2024Kuftinova} explores the new reality of DT based on a transportation network model to manage the data of transportation infrastructure with LLM. However, the challenges of complex system, incomplete knowledge and so on require new strategies. \cite{Manias2024} describes the LLMs developed for intent-based networks and explores how future LLMs can enable fully automated network intelligence and end-to-end networking.

In conclusion, complex network demands for more intelligent resource allocation strategies and more automated network management urgently. With the power of LLM, the integration of LLM and network management makes sense and is the general trend.

Our key contributions are as follows: 1) we propose a comprehensive LLM-based autonomous network management framework for DT application in 6G networks. The proposed framework consists of two stages: data-planing and modelling data retrieval as optimization problem with LLM; 2) we propose a full toolset including a cellular traffic load predictor, a parameter convertor and optimization solvers that can be used as LLM Function Calling to build efficient radio resource management (RRM) scheme automatically. In this case, we propose a novel and efficient network self-management technique tailored for complex network management scenarios in DT.

The remainder of the paper is organized as follows: \cref{sec:sys-model-prob-form} presents the system model and problem formulation. In \cref{sec:solution}, we detail our proposed solution and the implementation of the LLM-supported networks management framework. \cref{sec:sim-results} discusses the simulation results and performance evaluation. Finally, \cref{sec:conclusion} concludes the paper.

\section{System Model and Problem Formulation}
\label{sec:sys-model-prob-form}

In this section, we introduce the system model and formulate the resource management problem in the context of LLM-supported 6G-empowered DT networks management.

\subsection{System Model}
\label{sec:system-model}

\begin{figure}[t]
  \centering
  \includegraphics[width=\linewidth]{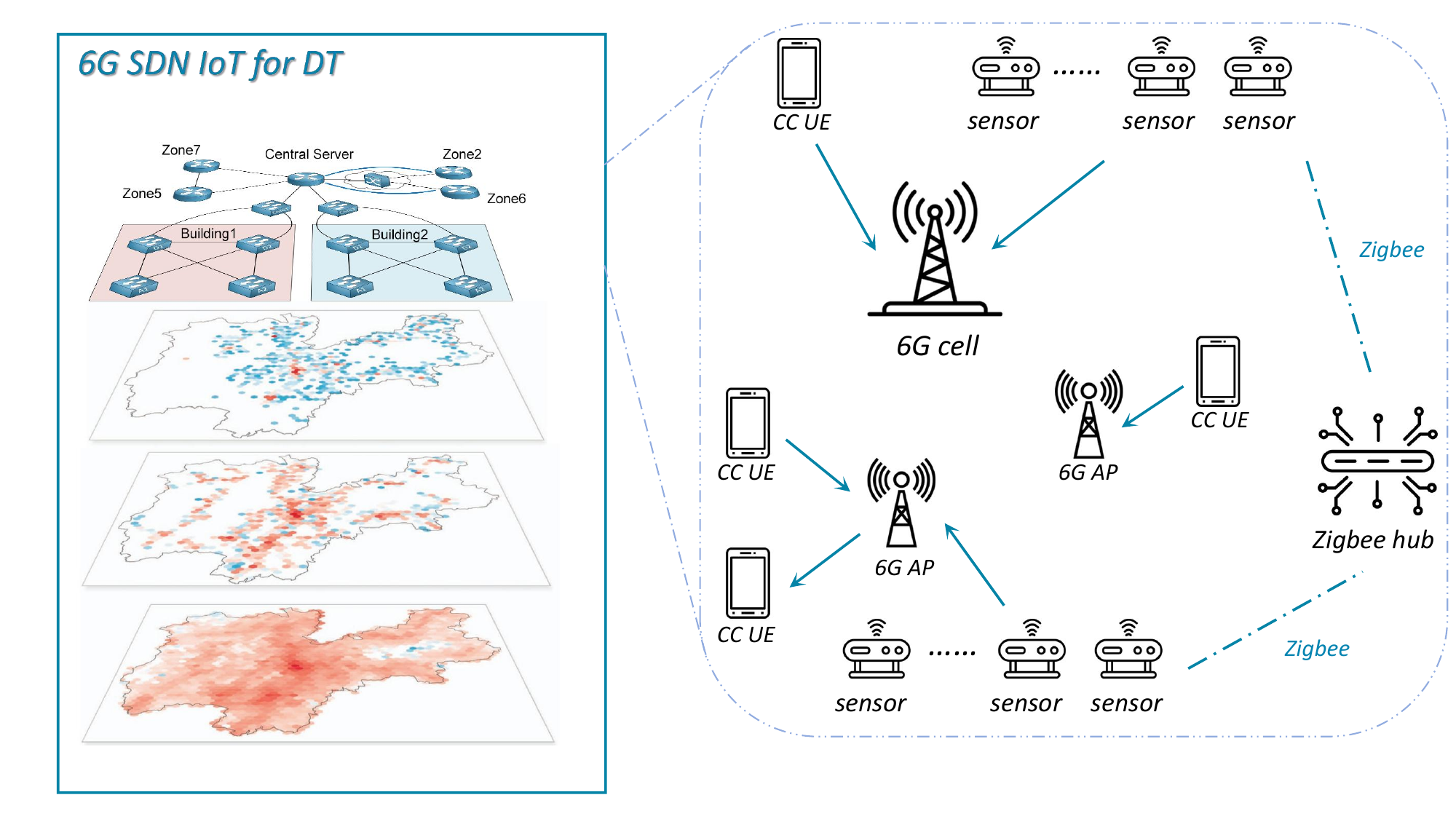}
  \caption{System model of the proposed LLM-supported network management framework for 6G-empowered DT networks.}
  \label{fig:system_model}
\end{figure}

We consider a 6G-enabled IoT network for DT, where IoT sensors collect data that is processed and managed by a centralized server using LLMs for intelligent management, as depicted in \cref{fig:system_model}.

We assume that IoT sensors are distributed throughout the city. We utilize SDN technology in the 6G network to manage communications between sensors and the central node~\cite{6g-sdn}. As shown in~\cref{fig:system_model}, the control plane of SDN can be divided into different layers. The top layer is the logic connections among sensors and routers, and the bottom layers consists of network traffic load, geography distribution of sensors, IoT application data load, which can be used to decide the routing strategy of SDN network. For the physical sensor network, the sensors will be connected to the 6G network if the traffic load is lower than a threshold $\tau$. Otherwise, it will fall back to Zigbee connections~\cite{Zigbee} (see in the right part of \cref{fig:system_model}). We consider a lazy-load characteristic of the sensors' data as described in~\cite{dt}. That is, the data will be stored at the sensors (edge storage) until they are actually required by the DT application and then transmitted to the central server for computation. In each cellular network, there exists conventional cellular (CC) UEs. 

In this paper, we assume that if the IoT communications in Zigbee mode will follow the default Zigbee settings, and we mainly consider the RRM problem in 6G networks. We consider the uplink orthogonal multiple access (OMA) system in sub-5GHz band for Macro Urban communications as described in~\cite{3gpp}. The channel model for IoT communications is expressed as:
\begin{equation}
  \label{eq:channel-model}
  g = \kappa d^{-\alpha} \|h\|^2 \zeta,
\end{equation}
where \( \kappa \) is an environment-related constant, \( \alpha \) is the path loss exponent, \( d \) is the distance between the transmitter and receiver, \( h \) is the Rayleigh fading coefficient, and \( \zeta \) denotes the log-normal distributed shadowing. The signal-to-interference-plus-noise ratio (SINR) for the transmission of an IoT device \( i \) is:
\begin{equation}
  \label{eq:sinr-iot-bs}
  \gamma_{i,j} = \frac{g_{i,j} p_i}{\sum_{k \in N} g_{k,i} p_k + N_0},
\end{equation}
where \( g_{i,j} \) is the transmission channel gain of IoT communication, \( p_i \) is the transmit power of IoT device \( i \), $\sum_{k \in N} g_{k,i} p_k$ is the inter-cell interference and \( N_0 \) is the additive white Gaussian noise (AWGN) power.

\subsection{Problem Formulation}
\label{sec:prob-formulation}

The objective is to minimize the largest delay of data transmission from the IoT devices to the BS while maintaining the QoS requirement of all UEs.

Let \( \bm{p} = [p_1, p_2, \ldots, p_N] \) denotes the transmit power vector for the \( N \) IoT devices. $D_i$ is the data size of IoT device $i$ to be transmitted to the BS. The optimization problem can be formulated as follows:


\begin{equation}
  \label{eq:objective-case1}
  \text{\emph{OPT}: } \min_{\bm{p},\bm{b}}\max \frac{D_i}{\sum b_{ij} R_{ij}},
\end{equation}
subject to:
\begin{align}
  &0 \leq p_i \leq p_{\max}, \quad \forall i \in \{1, 2, \ldots, N\}, \label{eq:power-constraint-case1} \\
  &\gamma_{i,bs} \geq \beta, \quad \forall i \in \{1, 2, \ldots, N\}, \label{eq:sinr-constraint-case1} \\
  & b_{ij} = \{0, 1\}
\end{align}
where \( R_{ij} \) is the achievable data rate of IoT device \( i \) on resource block (RB) $j$ given by:
\begin{equation}
  \label{eq:data-rate-case1}
  R_{ij} = B \log_2 (1 + \gamma_{ij}),
\end{equation}
\( p_{\max} \) is the maximum transmit power of an IoT device, and \( \beta \) is the SINR threshold required for reliable communication. $b_{ij}$ is the indicator function that represents whether an RB is allocated to the IoT communication.

The constraints in \eqref{eq:power-constraint-case1} ensure that the transmit power of each IoT device remains within the permissible range, while the constraints in \eqref{eq:sinr-constraint-case1} ensure that the SINR for each IoT device meets the required threshold for reliable data transmission.

\section{LLM Agentic Workflow for Network Management}
\label{sec:solution}

\begin{figure}[t]
  \centering
  \includegraphics[width=\linewidth]{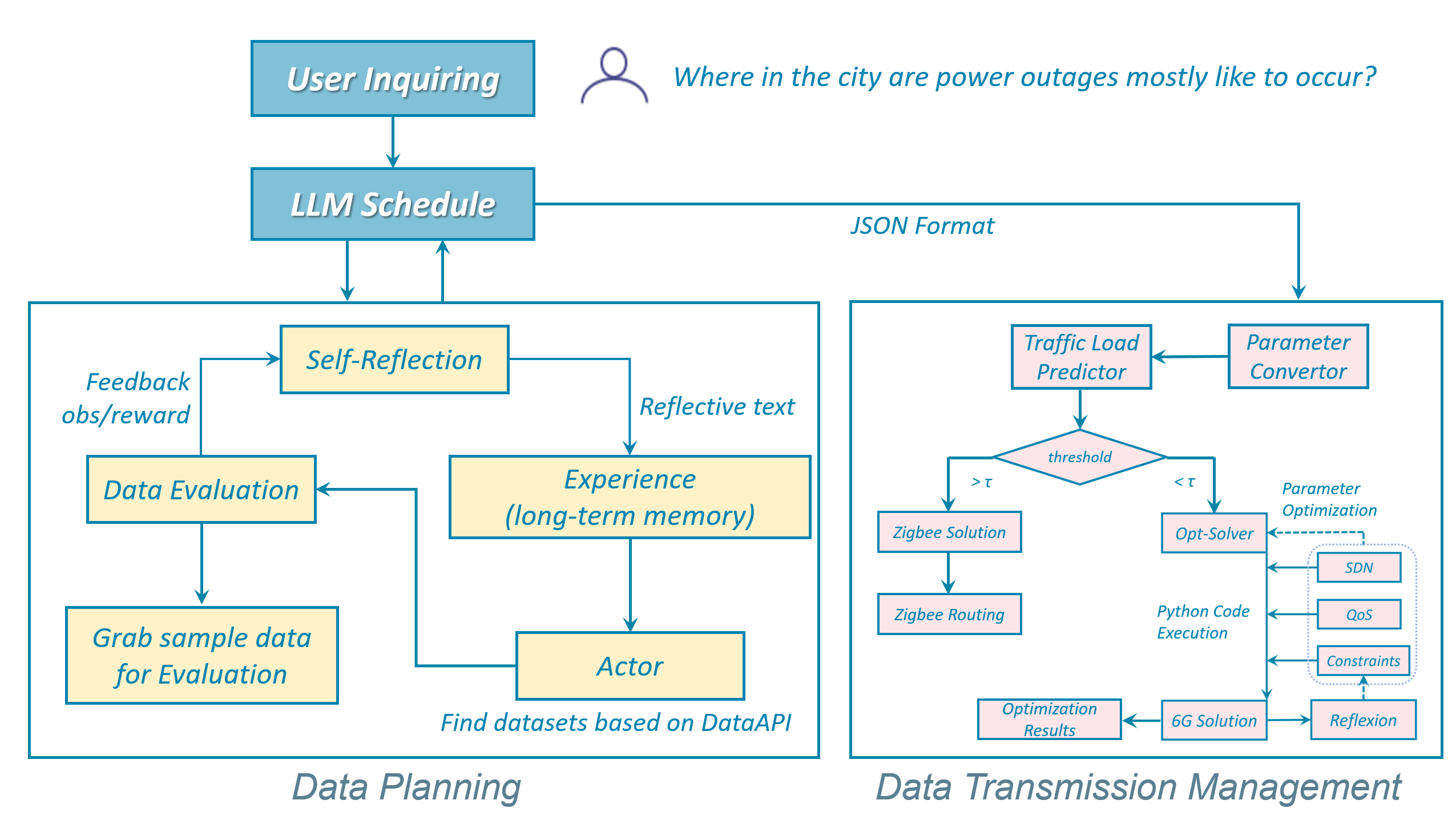}
  \caption{The overall workflow for LLM-based DT Management System}
  \label{fig:llm-flow}
\end{figure}

\begin{figure}[t]
  \centering
  \includegraphics[width=\linewidth]{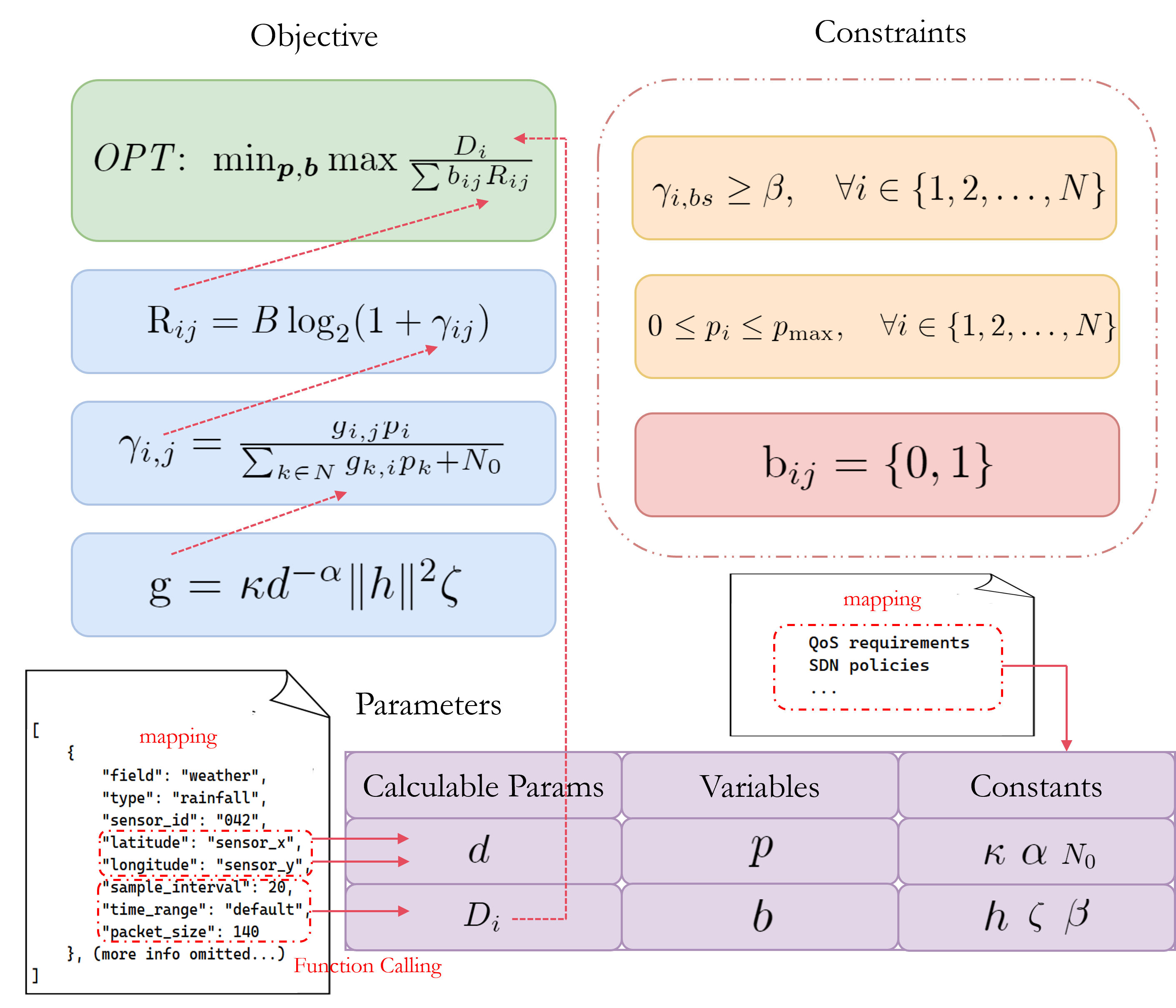}
  \caption{The detailed working principle of \emph{Parameter Convertor}}
  \label{fig:convertor}
\end{figure}

In this section, we propose LINKs, an agentic workflow for addressing data retrieval and RRM. LINKs consists of two main stages. The first stage centered around LLM-based scheduling serves as data planning, where LINKs identifies all necessary data based on the user query and returns relevant results by utilizing multi-agent framework. The second stage, on the other hand, focuses on RRM and data transmission where there is an emphasis on a series of tool use and reflexion mechanism that brings about automated parameter tuning to solve the optimization problem until all constraints are satisfied. An overview of the workflow of LINKs is depicted in \cref{fig:llm-flow}.

\begin{figure*}[t]
  \centering
  \includegraphics[width=\linewidth]{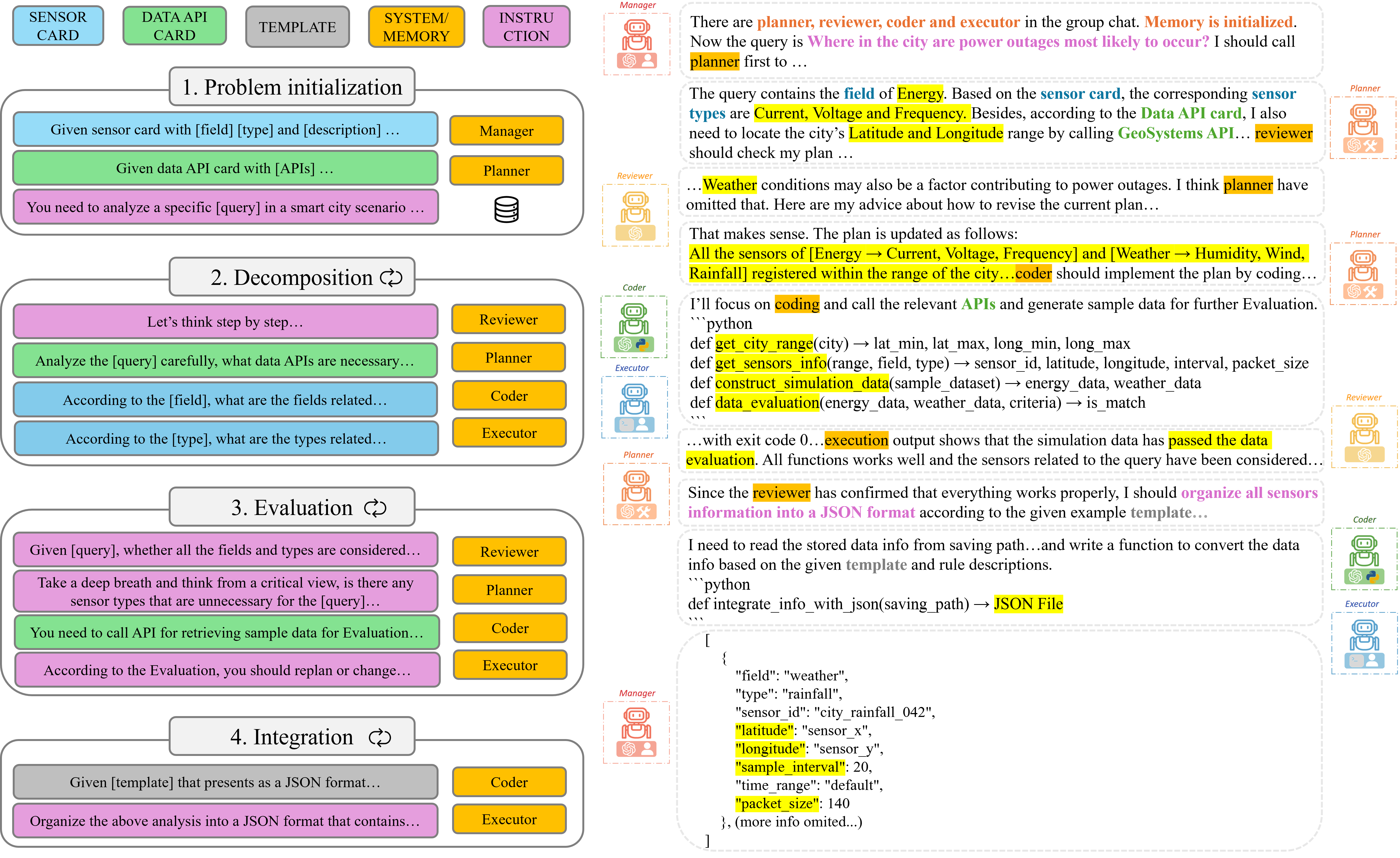}
  \caption{Prompt design for \emph{Data Planning} stage. The first stage of LINKs can be divided into four sub-parts from top to bottom, where, aside from initialization, the loop symbols within each sub-part denote potential iterations in the dialogue. On the right side, critical analysis and conclusions for example query in the chat are presented. Intermediate results are highlighted to clarify the outputs at each step of the first stage.}
  \label{fig:first-stage}
\end{figure*}

\subsection{Data Retrieval and Radio Resource Management}
\label{sec:data-planing}
Our multi-agent framework and demonstration of prompt engineering is illustrated in~\cref{fig:first-stage}.

We firstly develop \emph{Sensor Card} and \emph{Data API Card} to describe IoT sensors and data in a unified manner.

\emph{Sensor Card} is consist of three distinct fields, each specifying several particular sensor types and their standardized descriptions. For example, Temperature sensor is defined as follows: 

\textit{It is designed to monitor ambient temperature, utilizing degrees Celsius as the standardized unit of data representation and storage.}

\emph{Data API Card} incorporates predefined API functions crucial for data-related tasks, including mapping location entities to geographic coordinates, and searching for registered sensor information that maintained in the database.

LINKs draws inspiration from ReAct \cite{react} and Self-Reflection \cite{reflexion}, integrated within a multi-agent framework\cite{wu2023autogen}. LINKs has a \textbf{Manager} that directly links to other agents (\textbf{Planner}, \textbf{Reviewer}, \textbf{Coder} and \textbf{Executor}). At each step, \textbf{Manager} reviews the conversation so far and chooses the next agent to advance the problem. Through communication and collaboration among these agents, LINKs can address various user queries under city contexts, such as traffic prediction, weather conditions, power distribution leakage analysis, tourism recommendations and etc.

User queries start at the Actor phase, where \textbf{Planner} analyzes and interprets them, identifying possible geographical entities and temporal expressions. \textbf{Planner} then devises a strategy to call available APIs. Most importantly, relevant sensor types in relation to the query are carefully considered.

During the Evaluation and Self-Reflection phase, \textbf{Reviewer} checks the integrity and accuracy of the whole plan, identifies potential errors and offers modification suggestions. Reflective texts are then integrated into long-term memory, guiding subsequent actions of the Actor in an iterative cycle.

When \textbf{Coder} is called, it initially interprets the given plan and generates code for \textbf{Executor} to either run or debug. The inner loop between \textbf{Coder} and \textbf{Executor} ensures the syntactical accuracy of the code, while the outer loop between \textbf{Coder} and \textbf{Reviewer} guarantees the logical correctness. These agents collaborate to ensure all functional code runs smoothly.

As shown in~\cref{fig:first-stage}, LINKs will output a JSON file containing all the data to be retrieved, and then LINKs should call \emph{Parameter Convertor} to map the sensor information in JSON format into calculatable parameters, as shown in \cref{fig:convertor}. Subsequently, \emph{Traffic Load Predictor}, as discussed in~\cref{sec:tl-prediction} is called to assess whether the traffic load is within the threshold $\tau$. If the traffic load exceed $\tau$, then the problem will be simply transformed into Zigbee routing. Otherwise, opt-solver will be called to address the optimization problem defined in (\ref{eq:objective-case1}). In this paper, we utilize Python as the programming language and Gurobi and Pyomo as the solvers. By incorporating the reflexion stage, the previous solution will be evaluated against the constraints one by one, guiding LINKs to tune the parameters until all constraints are satisfied. Ultimately, this process will yield the optimization results for data retrieval.

\subsection{Cellular Traffic Load Prediction}
\label{sec:tl-prediction}
We utilize the large foundation model for time-series forecasting model TimesFM proposed in~\cite{das2024decoderonlyfoundationmodeltimeseries} for cellular traffic load prediction. For TimesFM, the input patch contains 32 time points, while the output (prediction) is 128 time points. We re-scale the data set to meet this requirement (e.g., to predict 600 minutes traffic load, we calculate the traffic load for duration of 5 minutes). Two experiments are carried out in this paper: 1) zero-shot learning, which applies TimesFM to the traffic load dataset directly~\cite{das2024decoderonlyfoundationmodeltimeseries}; 2) Fine-tuning with LoRA~\cite{lora}. 

\section{Simulation Results}
\label{sec:sim-results}

In this section, we present the simulation results to evaluate the performance of our proposed agentic workflow in the context of DT applications in 6G networks. We simulate 10 CUEs, 10 Zigbee Coordinators, and adopt the Urban Macro (UMa) scenario and the channel model with frequency at 6GHz, as described in Sections 7.2--7.5 in~\cite{3gpp}. We utilize the open datasets described in \cite{milan-multi}, which consists of the distribution of IoT devices, the historical telecommunications load, power grid usage, weather and etc, of Milan and Trentino. We reorganize and constructe the descriptions for the sensors based on the dataset's format and use the dataset to evaluate the performance of our proposed framework.

We create a dataset comprising a collection of queries to evaluate the performance of our solution. Among them, three involve a single field, five encompass two fields, and the remaining two encompass all three fields. The complexity of the queries ranges from low to high, as illustrated in \cref{tab:query-dataset}.

\begin{table}[t]
\centering
\caption{The comparison of Accuracy results}
\label{tab:query-dataset}
\begin{tabular}{p{0.2cm}p{4.4cm}p{0.4cm}p{0.6cm}p{0.6cm}}
\toprule
\textbf{} & \textbf{Query} & \textbf{Field} & \multicolumn{2}{c}{\textbf{Accuracy}} \\
\cmidrule(lr){4-5}
 & & & \textbf{Naive} & \textbf{Ours} \\
\midrule
\midrule
Q1 & What is the average temperature of the city over the past two months? & 1 & 0.96 & 1 (+0.04) \\
\midrule
Q2 & How does the humidity level vary across different districts of the city today? & 1 & 0.92 & 1 (+0.08) \\
\midrule
Q3 & Which areas are at higher risk of flooding due to heavy rainfall this week? & 1 & 0.76 & 1 (+0.24) \\ 
\midrule
\midrule
Q4 & What is the correlation between power consumption and temperature fluctuations in residential areas? & 2 & 0.73 & 1 (+0.27) \\ 
\midrule
Q5 & What is the predicted impact of wind speed on the stability of power lines in the coastal areas? & 2 & 0.74 & 1 (+0.26) \\ 
\midrule
Q6 & How does the frequency of the power grid change with varying solar energy input throughout the day? & 2 & 0.64 & 0.94 (+0.30) \\ 
\midrule
Q7 & Where in the city are power outages most likely to occur? & 2 & 0.63 & 0.9 (+0.27) \\ 
\midrule
Q8 & What is the likelihood of electrical leakage in the city due to continuous rainfall? & 2 & 0.59 & 1 (+0.41) \\ 
\midrule
\midrule
Q9 & I plan to drive my Tesla to the park for an open-air barbecue tomorrow, so I need to charge it beforehand. What would be the best time to depart? & 3 & 0.60 & 0.86 (+0.26) \\ 
\midrule
Q10 & What is the probability of traffic accidents and power distribution leaks occurring simultaneously due to weather conditions in the city in one day? & 3 & 0.48 & 0.79 (+0.31) \\ 
\bottomrule
\end{tabular}
\end{table}

We use F1-score to evaluate the performance of the first stage of LINKs:

\begin{equation}
\text{\emph{Accuracy}}=\frac{2 \times \mathcal{P} \times \mathcal{R}}{\mathcal{P} + \mathcal{R}}
\end{equation}
where $\mathcal{P}$ is the precision and $\mathcal{R}$ is the recall.

During the accuracy comparison experiment, \textbf{Ours} method is applied with the configuration where model is gpt-4o, temperature is set to 0.1 and human-in-the-loop option is NEVER, while \textbf{Naive} method refers to a zero-shot approach using gpt-4o.
The simulation results presented in \cref{tab:query-dataset} indicate that our solution demonstrated strong performance in modeling smart city issues compared with just using a performative language model, achieving an average accuracy of 0.95 across a set of 10 questions of varying difficulty. Furthermore, as the difficulty increases, the accuracy improvement of our method becomes more significant, with the maximum performance gain reaching 41 percent (see Q8).

\begin{table}[t]
\centering
\caption{Ablation Experiment Results for various Models and Temperature settings}
\label{tab:experiment_results_q1}
\begin{tabular}{p{3cm}p{2cm}p{2cm}}
\toprule
\textbf{Model} & \textbf{Temperature} & \textbf{Accuracy} \\ 
\midrule
\multirow{3}{*}{gpt-4o} & 0.1 & 1 \\ 
 & 0.5 & 1 \\ 
 & 0.9 & 0.95 \\ 
\midrule
\multirow{3}{*}{gpt-3.5-turbo} & 0.1 & 1 \\ 
 & 0.5 & 0.95 \\ 
 & 0.9 & 0.9 \\ 
\midrule
\multirow{3}{*}{gemini-1.5-pro} & 0.1 & 0.86 \\ 
 & 0.5 & 0.81 \\ 
 & 0.9 & 0.8 \\ 
\midrule
\multirow{3}{*}{Llama-3.1-8B-Instruct} & 0.1 & 0.91 \\ 
 & 0.5 & 0.86 \\ 
 & 0.9 & 0.8 \\ 
\midrule
\multirow{3}{*}{Qwen2-7B-Instruct} & 0.1 & 0.91 \\ 
 & 0.5 & 0.91 \\ 
 & 0.9 & 0.81 \\
\bottomrule
\end{tabular}
\end{table}

Table~\ref{tab:experiment_results_q1} are ablation studies. For the same model with different temperature, the results show that accuracy slightly decreases as the temperature increases. It can be inferred that a lower temperature ensures more stability in the agentic design pattern, whereas a higher temperature often brings diversity of outputs, thus leading to more over-selections and penalties on accuracy. 

Sometimes a prolonged chat is unacceptable due to the resource limitations and latency requirements for the centralized server in the DT networks. Therefore, in the second ablation study, the steps of the workflow is limited.

\begin{figure}[t]
  \centering
  \includegraphics[width=\linewidth]{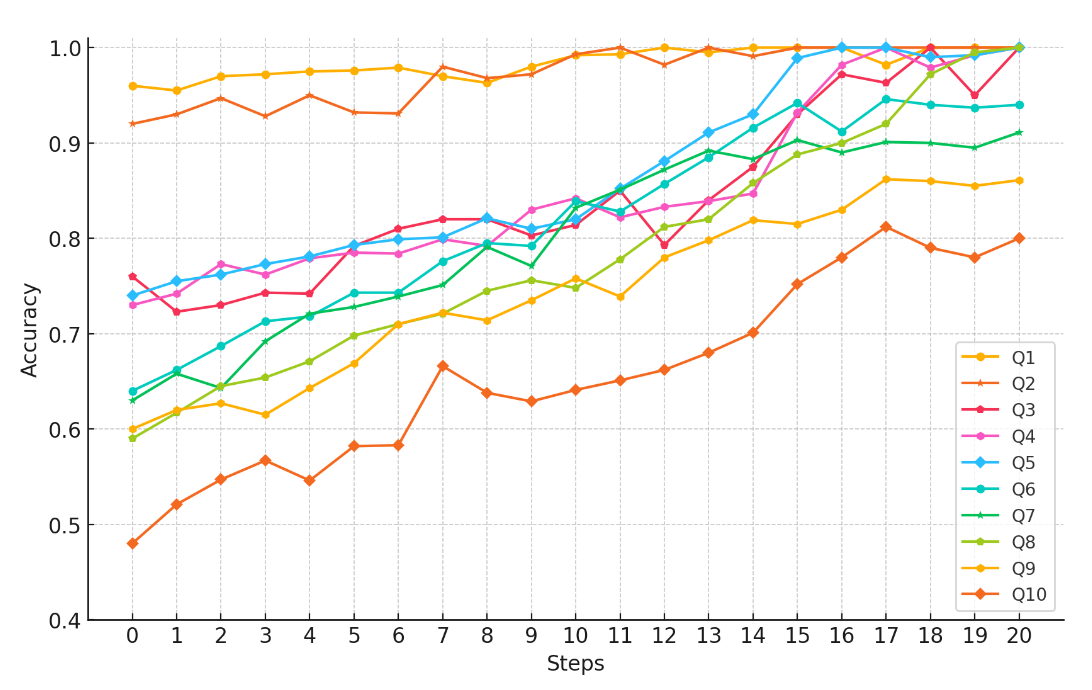}\label{fig:acc-step}
  \caption{Accuracy variation trends with the number of Steps}
  \label{fig:accuracy-step}
\end{figure}

Simple queries like Q1 and Q2 show nearly horizontal accuracy trends because they are rather straightforward. For more complex queries, accuracy generally increases with the number of steps and then gradually levels off after a certain point. As shown in \cref{fig:acc-step}, the critical point is around 15 steps. We conclude that an effective strategy is to choose gpt-4o with a low temperature in LINKs configuration settings. Additionally, it is worth noting that applying appropriate step limitations can help achieve a balance between performance and efficiency.

We then evaluate the performance of the proposed \emph{Traffic Load Predictor}. In this paper, NRMSE is used to evaluate the experimental results.
\begin{equation}
    \text{NRMSE} = \frac{1}{\bar{d}} \sqrt{\frac{1}{N} \sum_{k=1}^{N} \left( \hat{d}_k - d_k \right)^2}
\end{equation}
where \(\hat{d}_k\) represents the predicted value, \(d_k\) is the corresponding true value, \(N\) denotes the total number of measurement points in space and time, and \(\bar{d}\) is their average value. Since NRMSE eliminates scale dependence, it is commonly used for comparison between datasets or models of different scales. The smaller the NRMSE, the more accurate the model's prediction. It is a widely used evaluation metric in current research on wireless network traffic prediction problems.

\begin{table*}[t]
\caption{NRMSE of different models}
\label{tab:NRMSE}
\centering
\begin{tabular*}{\linewidth}{@{\extracolsep{\fill}} cccccccc}
\toprule[2pt]
\diagbox{Time}{Model} &\textbf{MLP} &\textbf{ARIMA} &\textbf{LSTM} &\textbf{3D-Conv} &\textbf{ConvLSTM} &\textbf{TimesFM} & \textbf{TimesFM+LoRA}\\
\midrule
1-step (10 minutes) & 0.5269$\pm$0.2257 & 0.6360$\pm$0.2078 & 0.5473$\pm$0.0879 & 0.4617$\pm$0.5033 & 0.4812$\pm$0.0576 & 0.3981$\pm$0.4110 & 0.3035$\pm$0.3290\\
10-step (100 minutes) & 0.5794$\pm$0.0929 & 0.6416$\pm$0.3655 & 0.5623$\pm$0.2741 & 0.4783$\pm$0.5006 & 0.5194$\pm$0.1146 & 0.4129$\pm$0.4118 & 0.3184$\pm$0.3364\\
30-step (300 minutes) & 0.6604$\pm$0.2001 & 0.6562$\pm$0.1618 & 0.6154$\pm$0.2174 & 0.5290$\pm$0.5884 & 0.5373$\pm$0.2025 & 0.4319$\pm$0.4383 & 0.3373$\pm$0.3571\\
60-step (600 minutes) & 0.7080$\pm$0.2424 & 0.7177$\pm$0.1200 & 0.6779$\pm$0.3110 & 0.6560$\pm$0.6983 & 0.5847$\pm$0.2018 & 0.4738$\pm$0.5311 & 0.3541$\pm$0.3909\\
\bottomrule[2pt]
\end{tabular*}
\end{table*}

We compare the zero-shot TimesFM and TimesFM with LoRA with some state-of-the-art traffic load prediction algorithms proposed in previous studies~\cite{10.1145/3209582.3209606} (see in \cref{tab:NRMSE}). We use the Huggingface PEFT library to perform the LoRA training, with paremeters set in \emph{peft.LoRAConfig} as: rank $r=16$, scaling factor lora\_alpha $=16$, lora\_dropout $=0.1$, bias $=none$, and targets the query and value matrices of the attention blocks. We set the batch size as 128 and the number of epochs as 10. From \cref{tab:NRMSE}, we conclude that even with zero-shot TimesFM, it significantly outperforms the SOTA algorithms. With LoRA, we can further improve the prediction accuracy.

\begin{figure}[t]
    \centering
    \includegraphics[width=\linewidth]{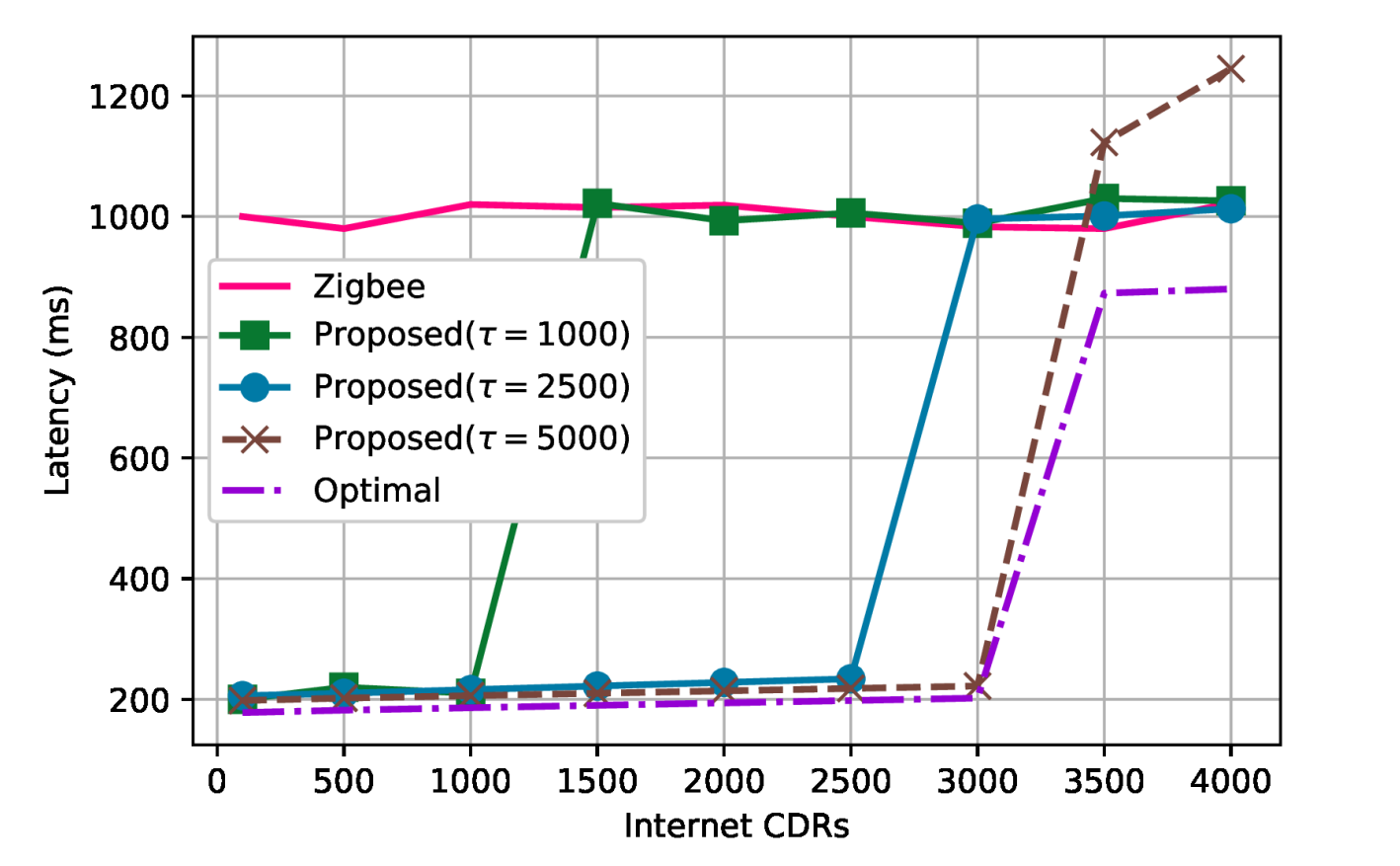}
    \caption{Latency performance of the proposed framework}
    \label{fig:latency}
\end{figure}

Finally, the overall system performance is evaluated in~\cref{fig:latency}. We conclude that the network latency of the proposed mechanism is significantly affected by $\tau$. For a specific $\tau$, with the high accuracy of the proposed LLM-based selection of data required and optimization problem solving, LINKs can achieve a near optimal solution.

\section{Conclusion}
\label{sec:conclusion}

We develop LINKs, an LLM-based management framework for DT applications in 6G NetworKs, that focuses on optimizing data retrieval and communication efficiency in smart city scenarios. During the \emph{Data Planning} stage, multi-agent framework ensures the accuracy of sensor data needed. The \emph{Data Transmission Management} stage models RRM as an optimization problem, using function calling to solve it autonomously in a clear pipeline, ultimately retrieving the required data with minimal latency.

According to simulation results, LINKs exhibits robust performance in modeling smart city issues, achieving extremely high accuracy for the given dataset, showing even greater advantages in more challenging queries. We use Google's TimesFM, a pretrained time series prediction model, and further finetuned it with LoRA as the \emph{Traffic Load Predictor} to achieve more accurate time prediction. Finally, in the overall system latency performance evaluation, LINKs can achieve a near-optimal solution, again showcasing its superior performance.



\bibliographystyle{IEEEtran}
\bibliography{6gdt.bib}

\end{document}